\documentclass[12pt]{iopart}
\usepackage{epsfig}
\usepackage{amssymb}
\usepackage{rotating}
\usepackage{pdflscape}
\begin{document}

\title[How do I introduce Schr\"odinger equation]{
  How do I introduce Schr\"odinger equation during the quantum mechanics course?}

\author{T. Mart}

\address{Departemen Fisika, FMIPA, Universitas Indonesia, Depok 16424, Indonesia}
\ead{terry.mart@sci.ui.ac.id}
\vspace{10pt}
\begin{indented}
\item[]
\end{indented}

\begin{abstract}
  In this paper I explain how I usually introduce the Schr\"odinger equation
  during the quantum mechanics course. My preferred method is the chronological
  one. Since the Schr\"odinger equation belongs to a special case of 
  wave equations I start the course with introducing the wave equation. The
  Schr\"odinger equation is derived with the help of the two quantum concepts
  introduced by Max Planck, Einstein, and de Broglie, i.e., the energy of a
  photon $E=\hbar\omega$ and the wavelength of the de Broglie wave
  $\lambda=h/p$. Finally, the difference between the classical
  wave equation and the quantum Schr\"odinger one is explained in order 
  to help the students
  to grasp the meaning of quantum wavefunction $\Psi({\bf r},t)$.
  A comparison of the present method to the approaches given by
  the authors of quantum mechanics textbooks as well as that of
  the original Nuffield A level is presented. It is found that the 
  present approach is different from those given by these authors, except
  by Weinberg or Dicke and Wittke. However, the approach is in 
  line with the original Nuffield A level one.

\end{abstract}

\begin{indented}
\item[]\today
\end{indented}

\section{Introduction}
\label{sec:introduction}
Quantum mechanics is a notoriously difficult and abstract topic. I remember that
when I started my first year of my undergraduate study more than 30 years 
ago, my friend told me that in the following years we were going to hear about 
Schr\"odinger equation which is elegant but complicated and difficult. I was 
very excited at that time and felt that I could not wait any longer to attend 
the quantum mechanics course. However, before I could acquaint this 
Schr\"odinger masterpiece, I accidentally entered the Introduction to Solid
State Physics course, where I had to calculate the probability of an electron 
transition represented by the Dirac bracket that sandwiches a potential 
operator between two wavefunctions. Immediately, I got the opinion that 
quantum mechanics was very difficult, not interesting,
and I had even no idea why should people write such a bracket. 

It is not an exaggeration if Feynman once said in his well known quote:
''I think I can safely say that nobody understands quantum mechanics
\cite{feynman_the_character}.'' Of course what Feynman really means is
written a few paragraphs before the quote, i.e., the quantum-mechanical
objects behave in a way that we have never seen or heard before. The 
objects are governed by different law, different from what we experience 
everyday.

However, things changed dramatically 
after I took quantum mechanics courses at different
levels and started to really use it in my Ph.D. research. Now, after
more than 20 years teaching quantum mechanics in my university
I notice a number of explanations that were missing during the 
quantum mechanics courses I attended before. One of them is the 
clear and efficient method to introduce the Schr\"odinger equation. 
It is sometimes
forgotten that the introduction of Schr\"odinger equation during
the quantum mechanics course is actually a critical step, since it is 
the time to change the students' classical-mechanics thinking to 
the quantum-mechanics one. 

The following discussion is not only relevant to the course of quantum
mechanics, but also important in the modern physics course, where the
matter wave concept is introduced. Indeed, nowadays, 
introducing the quantum mechanics concept is already important 
at the high school level \cite{henriksen,michelini}. 

\section{Introducing the classical wave equation}
\label{sec:classical}

\begin{figure}[!h]
  \begin{center}
    \leavevmode
    \epsfig{figure=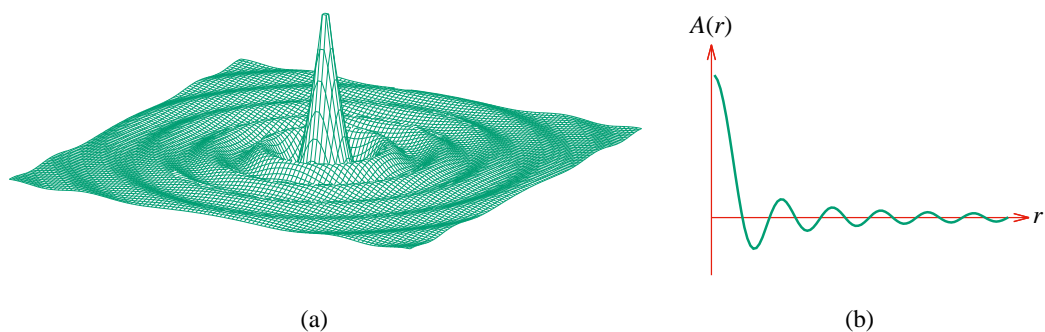,width=140mm}
    \caption{(a) Example of a very intuitive wave phenomenon,
             but mathematically a slightly complicated one. (b) The one-dimensional
             cut of this wave shows that the oscillation of this wave is damped.}
   \label{fig:water-waves}
  \end{center}
\end{figure}

Since the Schr\"odinger equation is a differential wave equation, 
it is essential to remind the students about the wave equation 
in the classical mechanics. However, it should be remembered that 
during that time not all students can quickly understand the wave
equation in terms of sinusoidal function or differential equation. 
Therefore, I usually start with the very intuitive wave phenomenon 
depicted in Fig.~\ref{fig:water-waves}(a). This phenomenon can be 
observed if we, e.g., drop a stone in the water. Since the propagation 
of the wave is radial and the amplitude decreases as the radius 
increases, the simplest stationary wavefunction reads
\begin{equation}
  \label{eq:water_waves}
  A(r) = A_0\,\frac{\sin r}{r} ~,
\end{equation}
where $A(r)$ is the displacement of the water at the position $r$ and 
$A_0$ indicates the amplitude at $r=0$. Note that the wavefunction
of the water ripples is actually not stationary, it is a function of time and
although we only consider the stationary case given by Eq.~(\ref{eq:water_waves})
it is still not simple. It has a dependence on $r$ in the denominator which is
indeed important to suppress the amplitude in the location far from the wave 
center, as clearly shown by Fig.~\ref{fig:water-waves}(b).

\begin{figure}
  \begin{center}
    \leavevmode
    \epsfig{figure=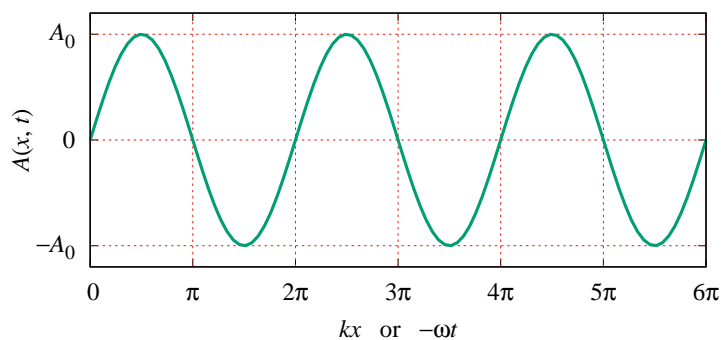,width=100mm}
    \caption{A mathematically simple wave phenomenon, but less intuitive
      one, i.e., $A(x,t)= A_0\,\sin(kx-\omega t)$.}
   \label{fig:simple-waves}
  \end{center}
\end{figure}

A mathematically much simpler wave equation is the sine wave displayed in 
Fig.~\ref{fig:simple-waves}. Example of the sine wave is the standing wave
observed in the string of a guitar. However, since everybody can 
easily drop a
stone in the water, whereas not everybody has an access to a guitar, the
second example is less intuitive. Nevertheless, all students have learned standing
wave in high school and since the wave is very simple, we will start
our calculation with the sine wave. 

The wavefunction displayed in Fig.~\ref{fig:simple-waves} can be written as
\begin{equation}
  \label{eq:sine_wave}
  A(x,t) = A_0\,\sin (kx-\omega t) ~,
\end{equation}
where $k$ is the wave number and $\omega$ is the angular frequency. A computer
software like {\it Mathematica} or {\it Matlab} can help the lecturer to 
simulate this wave during the course. Note that
the two parameters define the phase velocity of the wave, i.e.,
\begin{equation}
  \label{eq:velocity}
  v=\omega/k ~.
\end{equation}
By taking the second derivatives of Eq.~(\ref{eq:sine_wave}) with respect to
$x$ and $t$ and using Eq.~(\ref{eq:velocity}) it is easy to show that
\begin{equation}
  \label{eq:wave-equation}
  \frac{\partial^2 A(x,t)}{\partial t^2} = v^2\,
  \frac{\partial^2 A(x,t)}{\partial x^2} ~.
\end{equation}
Equation~(\ref{eq:wave-equation}) is the wave differential equation in 
classical mechanics.
All parameters and variables given in Eq.~(\ref{eq:wave-equation}) are real
and physical. For instance, the displacement $A(x,t)$ is real and can be 
observed. For the three dimensional problem this equation is given in
the form of the Laplacian operator  $\nabla^2$, i.e.,
\begin{equation}
  \label{eq:wave-equation-3D}
  \frac{\partial^2 A({\bf r},t)}{\partial t^2} = v^2\,
  \nabla^2\, A({\bf r},t) ~.
\end{equation}

Note that, since the cosine function can also describe this wave, it is also
possible to generalize the function by using the complex exponential one. 
It is assumed that students have been acquainted with the imaginary number 
$i=\sqrt{-1}$ during the calculus course. The general form of Eq.~(\ref{eq:sine_wave})
reads
\begin{equation}
  \label{eq:exp_wave}
  A(x,t) = A_0\,e^{i(kx-\omega t)} ~.
\end{equation}
The displacement $A(x,t)$ is still observable and real by keeping in mind that the
observed quantities are represented by the real and imaginary parts of 
$A(x,t)$. Here, the imaginary number $i$ works as a unit vector pointing 
to completely different direction from the real one. Thus, the $i$ number 
completely separates the sine and cosine waves.  

\section{Introducing quantum concepts to obtain the  Schr\"odinger equation}
\label{sec:quantum}
Before introducing the  Schr\"odinger equation it is important to explain
to students why Schr\"odinger needed this equation at that time. Therefore,
a brief chronological history of this equation is strongly advocated. 

We begin with the finding of Max Planck in 1900 which was then proven 
by Einstein in 1905 that the energy of electromagnetic wave is carried 
out by the quanta (packets) called photons. Each quantum carries energy 
of
\begin{equation}
  \label{eq:energy_quanta}
  E = h\, \nu ~=~ \hbar\,\omega ~,
\end{equation}
where $h$ and $\nu$ are the Planck constant and the frequency of 
the electromagnetic
wave, respectively, whereas $\hbar=h/2\pi$. Equation (\ref{eq:energy_quanta}) 
indicates the matter property of a wave, since by using the Einstein
energy momentum relation $E^2=p^2c^2+m_0^2c^4$ and remembering that the mass
of rest photon is zero, we obtain from Eq.~(\ref{eq:energy_quanta}) that
$pc= hc/\lambda$ or the (matter) momentum of the photon reads
\begin{equation}
  \label{eq:einstein}
  p = \frac{h}{\lambda} ~,
\end{equation}
where we have used the fact that the speed of electromagnetic wave 
$c=\lambda\nu$.

In 1924 Louis de Broglie had an idea that in addition to the photon theory of
Max Planck and Einstein all particles with momentum $p$,  including the massive
electrons, have also a wave property with the wavelength of
\begin{equation}
  \label{eq:de_broglie}
  \lambda = \frac{h}{p} ~.
\end{equation}
At first glance Eq.~(\ref{eq:de_broglie}) is nothing but 
Eq.~(\ref{eq:einstein}). Nevertheless, de Broglie wrote  
a 73 pages thesis to defense this idea \cite{debroglie}. 
Even the thesis committee did not know what to to with the 
thesis, so they sent it to Einstein. Fortunately Einstein 
saw the point and recommended the approval. Einstein sent 
the thesis to his colleagues he knew would be interested 
in the de Broglie's idea \cite{smolin}. 

On November 1925 Erwin Schr\"odinger attended a colloquium held by
his colleague 
Peter Debye at the E. T. H. Zurich. As noticed by Felix Bloch in his
talk given at the 1976 American Physical Society meeting \cite{bloch} 
he heard that during the colloquium 
Debye asked Schr\"odinger to seriously consider the thesis of
de Broglie. What Debye was looking for is actually the wave
equation which can properly describe the de Broglie's wave. 
As a Sommerfeld student Debye had learned that without wave equation one
cannot deal with waves properly. Thus,  de Broglie's thesis 
was lacked of this equation and Debye challenged Schr\"odinger to find it. 
Schr\"odinger took this challenge during his Christmas vacation and amazingly 
he was able to submit the result to  Annalen der Physik 
on 27th January 1926 \cite{schrodinger}.

Let us pretend to be Schr\"odinger and we accept the Debye's challenge 
to find the wave 
equation. By using Eq.~(\ref{eq:de_broglie}) and 
the relation between wave number and wavelength $k=2\pi/\lambda$, we obtain
\begin{equation}
  k = 2\pi\,\frac{p}{h} ~=~ \frac{p}{\hbar} ~.
\end{equation}
Thus, with the help of Eq.~(\ref{eq:einstein}) we can recast 
Eq.~(\ref{eq:exp_wave}) into the form
\begin{equation}
  \label{eq:exp_wave2}
    \Psi(x,t) = \Psi_0\,e^{i(px-E t)/\hbar} ~,
\end{equation}
where we have replaced the displacement $A(x,t)$ with $\Psi(x,t)$, along with
the amplitude, to indicate that we have included the quantum concept 
introduced by Max Planck, Einstein, and de Broglie, given by 
Eqs.~(\ref{eq:energy_quanta}) and (\ref{eq:de_broglie}), in the wavefunction.
In fact, Schr\"odinger introduced the $\Psi$ function in his equation 
and called it ``a new unknown function''
in his 1926 paper \cite{schrodinger}.

Now, since we are talking about a non-relativistic free particle 
described by a plane wave, the total energy of the particle 
is only kinetic energy, i.e., $E=p^2/2m$. By using this fact and 
taking the first derivative of the wavefunction given in
Eq.~(\ref{eq:exp_wave2}) with respect to $t$, as well as 
the second derivative with respect to $x$, we obtain
\begin{equation}
  \label{eq:schrodinger-eq}
  i\hbar\,\frac{\partial \Psi(x,t)}{\partial t} =
  -\frac{\hbar^2}{2m}\,\frac{\partial^2 \Psi(x,t)}{\partial x^2} ~,
\end{equation}
which is the wave equation for a free particle with mass $m$
we are looking for.
Since the right hand side of Eq.~(\ref{eq:schrodinger-eq}) corresponds to
the kinetic energy $T$, we may generalize this equation to describe a particle
moving under the influence of a potential energy $V$ and in three dimensional
coordinate system as
\begin{equation}
  \label{eq:schrodinger-general}
  i\hbar\,\frac{\partial \Psi({\bf r},t)}{\partial t} =
  -\frac{\hbar^2}{2m}\,\nabla^2 \Psi({\bf r},t) + V ({\bf r})\, \Psi({\bf r},t) ~,
\end{equation}
where we have used the fact that the total energy 
$E=T+V$. Equation 
(\ref{eq:schrodinger-general}) is the Schr\"odinger equation 
in general form. We note that similar derivations can be also
found in the literature, e.g., Refs.~\cite{dicke,gorard,chem,weinberg}.

\section{Comparing classical mechanics to quantum mechanics}
\label{sec:compare}
By introducing the Schr\"odinger equation in Section~\ref{sec:quantum} we 
have jumped to the quantum world which is very different from the classical
one. At this stage 
it is important to explain to the students the difference between the
classical wave equation given by Eq.~(\ref{eq:wave-equation-3D}) and its 
quantum counterpart given by Eq.~(\ref{eq:schrodinger-general}).

The first and very important difference is the ''displacement'' in the two
cases. In the classical case the displacement $A({\bf r},t)$ is physical
and observable. It is easy to imagine that the water ripples shown in 
Fig.~\ref{fig:water-waves}, or the string oscillation displayed in
Fig.~\ref{fig:simple-waves}, can be represented by the variation 
of $A({\bf r},t)$ due to the variations of {\bf r} and $t$. On the
contrary, $\Psi({\bf r},t)$ given in Eq.~(\ref{eq:schrodinger-general})
is not a displacement. It is just a wavefunction describing the de Broglie
wave given by Eq.~(\ref{eq:de_broglie}). 

Is the de Broglie wave physical? Obviously, the answer is no. 
The massive particles exhibit the wave phenomenon, but the wave 
itself is not physical. To comprehend 
this we have to go back to the experimental verification of 
Eq.~(\ref{eq:de_broglie}). The first experiment was performed by Davisson
and Germer in 1927 by using electron scattering on a single nickel crystal
\cite{davisson}. A more modern experiment using cold neutrons 
diffraction on single and double slits was performed by Zeilinger 
and his collaborators in Grenoble \cite{zeilinger}. 
In principle, it is similar to the conventional diffraction experiment. 
However, instead of using real photons (visible light) here one uses 
neutron beams and the general diffraction pattern can be reproduced.
Therefore, we may conclude that 
after passing the slits the neutrons exhibit one of the wave phenomena, 
i.e., diffraction. This clarifies that we do not detect the de Broglie 
wave, but we merely observe its phenomenon. 

As a consequence, the displacement  $A({\bf r},t)$ is real, whereas its 
counterpart $\Psi({\bf r},t)$ is allowed to be complex function. The latter
was actually problematic, even for Schr\"odinger himself, because he did not 
expect it. In 1926 Max Born came up with the idea that $\Psi({\bf r},t)$ is
a probability amplitude and its modulus squared 
is the probability density \cite{born}, which was considered by Schr\"odinger
with great doubt \cite{bloch}.

The second difference is, unlike the Schr\"odinger equation,
the wave  equation is symmetric in the order of derivatives.
This explains why the  Schr\"odinger equation is non-relativistic, 
since it cannot be formulated in a covariant form as in the case of
the wave equation for the photons. We know that the covariant formulation 
requires that
both space and time terms should be in the same order. 

\section{Comparison with other approaches}
To the best of my knowledge, there has been no explicit
discussion in the literature on the most effective method to 
introduce the Schr\"odinger equation during the course of 
quantum mechanics. However, since most of quantum mechanics 
textbooks are written based on the teaching experience of the 
authors, we can compare my approach explained here with the 
steps used by these authors before they write the Schr\"odinger 
equation. Furthermore, the Schr\"odinger equation is usually 
introduced in the first or second chapter of quantum mechanics 
textbooks. Thus, their approaches should be easily identified. 


Let us start with the book of Gasiorowicz \cite{gasiorowicz} 
which has been considered as one of the standard quantum mechanics 
textbooks used in most universities for relatively long time, 
i.e., since its first edition in 1974 up to now. In the Sakurai's 
book of Modern Quantum Mechanics the editor San Fu Tuan 
wrote that Gasiorowicz is well known as an 
effective teacher, who can make difficult concepts 
simple and clear even in advanced research areas of
particle theory \cite{sakurai}. In spite of this great 
testimony, however, Gasiorowicz introduces the Schr\"odinger 
equation abruptly in Chapter 2 in the form of differential 
equation, without an explanation why we need it. Fortunately, 
in Chapter 1 Gasiorowicz explains a number of phenomena that 
indicate the limits of classical physics or the need of quantum
mechanics in the beginning of 1900s. Included in these phenomena 
are the particle behaviour of waves found by Planck and Einstein, 
and the wave behaviour of particle found by de-Broglie. 
Nevertheless, since the historical background of quantum mechanics
is given in a separate chapter, Gasiorowicz's step is different
from my approach.

The books written by Griffiths are also very 
popular. Griffiths is well known through his clear and explicit 
derivations of the used formulas in all of his books. This can 
be also found in his quantum mechanics book \cite{griffiths}. 
However, in the case of the Schr\"odinger equation he also 
introduced it suddenly on page one and indeed without any 
historical background. Presumably, Griffiths emphasizes more 
mathematical formalism than chronological or historical 
approach. Thus, his approach is also different from mine.

Interestingly, a much older book written by Dicke and Wittke 
\cite{dicke} has a much closer approach to the one that I explained 
in Section \ref{sec:quantum}. 
Dicke and Wittke open the discussion in Chapter 1 with the 
evidence of the inadequacy of classical mechanics by 
explaining Max Planck theory of black-body radiation, 
theory of specific heats of solids according to Debye, 
Compton scattering, and their quantum-squirrel-cage 
gedanken experiment.  In Chapter 2 they introduce the de 
Broglie wave under the name of wave mechanics. In 
Chapter 3 these authors introduce the Schr\"odinger 
equation with a similar step as I explained in 
Section \ref{sec:quantum}, i.e., they start with 
Eq.~(\ref{eq:energy_quanta}) and finish with 
Eq.~(\ref{eq:schrodinger-general}). After introducing 
the Schr\"odinger equation Dicke and Wittke proceed 
to discuss its solutions for a number of one-dimensional 
potentials. 

More recent and modern quantum mechanics textbooks 
\cite{auletta,banks,binney,commins,mcintyre,basdevant,townsend} 
also introduce the Schr\"odinger equation abruptly and, 
in fact, it is mostly in the form of eigenvalue equation
$H\Psi=E\Psi$, with $H$ the Hamiltonian of the system. 
Although most of the books start with historical background 
of quantum mechanics, it is given in a separate chapter before 
the one where the Schr\"odinger equation is introduced. There seems to be 
an unwritten consensus among these authors that students should have 
already known the Schr\"odinger equation as in the case of 
classical mechanics, in which students have been already very 
familiar with the Newton laws. 

Among the recent quantum mechanics books, only the 
one  written  by Steven Weinberg is different 
\cite{weinberg}. Weinberg opens the book with the first 
chapter discussing the historical introduction, which 
is constructed in a chronological order. Indeed, in
the beginning of the first chapter  Weinberg wrote:
''The principles of quantum mechanics are so contrary 
to ordinary intuition that they can best be motivated 
by taking a look at their prehistory.''
Thus, in the Weinberg's book the Schr\"odinger equation 
is derived directly in Chapter 1, in a similar manner as 
the approach given in the present paper. 


Another approach which can be contrasted with the one explained
in this paper is the original Nuffield A level. We understand 
that the Nuffield A level is not intended for a rigorous explanation
of quantum mechanics. As a consequence, the derivation of 
Schr\"odinger equation is most probably beyond the topic of discussion
in the Nuffield A level.
However, the modern physics part of this approach has been 
extensively verified. Furthermore, as written by 
Fuller and Malvern \cite{fuller-malvern} the original Nuffield 
A Level Physics contains a unit that discusses the quantum mechanics 
explanation of the wave particle duality phenomenon seriously, i.e., 
in Unit 10: Waves, Particles and Atoms, which consists of
\cite{fuller-malvern} 
\begin{itemize}
\item photons, wave-particle duality,
\item electrons, electrons as a wave,
\item waves in boxes, Sch\"odinger's equation,
\item the scope of wave mechanics.
\end{itemize}
From the ordering of the four items above we may conclude that, 
in principle, the discussion of Sch\"odinger equation in the 
original Nuffield A level is in line
with my approach explained in this paper.

\section{Further consideration}
I agree with Steven Weinberg that the best way to introduce
quantum mechanics is by considering its chronological history
\cite{weinberg}.
This include the Schr\"odinger equation, which is the main
equation in non-relativistic quantum mechanics. Nevertheless, 
to obtain a more objective result a more quantitative investigation
should be performed in the class. This can be carried out in two
parallel student groups, where two different approaches can be
applied. Such an investigation is possible in my physics 
department, since each semester it runs two or three parallel
classes for the Introductory Quantum Mechanics course. Therefore,
it is my plan to study the effectiveness of the present
approach in the near future.

Having introduced the Schr\"odinger equation, the 
next step is presenting the applications of this equation,
i.e., variations of the potential $V$ in Eq.~(\ref{eq:schrodinger-general}). 
Of course, the best example is
the application for the Hydrogen atom as described in the 
Schr\"odinger paper \cite{schrodinger}. 
However, since simpler is better,
I like to start with the traditional approach, i.e., one-dimensional 
potential. My favorite case here is the simple potential barrier and
potential well, which have a wide range of applications, i.e., from scanning
tunneling microscope to the nuclear reactions in stars \cite{gasiorowicz}. 
Such applications have naturally strong impact on the student motivation to attend 
the course. 

\section{Summary and Conclusion}
\label{sec:conclusion}
I have discussed that introducing the Schr\"odinger equation during 
the quantum mechanics course is a critical step, since it constitutes 
a transition from classical mechanics to the quantum one. Therefore, 
a smooth but clear transition is required. For this purpose, I used 
a chronological step in the 
development of quantum mechanics. I introduced the classical wave 
equation prior to the quantum one. By using the quantum concepts of 
the photon energy and the de Broglie wavelength of a moving massive
particle I derived the Schr\"odinger equation following the step to derive the
classical wave equation. There are a number of fundamental differences between
the two equations, which must be highlighted  in order to understand
the Schr\"odinger one and the meaning of the wavefunction $\Psi$.
Although the approach explained in this paper is different from 
those given by most of the authors of quantum mechanics textbooks,
it is similar to the approach adopted by Dicke and Wittke, as well as
by Steven Weinberg.

\section*{Acknowledgment}
The author was partly supported by 
the 2020 PUTI Q2 Research Grant 
of Universitas Indonesia, under contract 
No. NKB-1652/UN2.RST/HKP.05.00/2020.


\section*{References}

\begin{thebibliography}{00}

\bibitem{feynman_the_character} 
  Feynman R 1965 {\it The Character of Physical Law} 
   (Cambridge: The MIT Press) p 129

\bibitem{henriksen}
  Henriksen E K \etal 2014 Phys. Educ. {\bf 49} 678

\bibitem{michelini}
  Michelini M \etal 2000 Phys. Educ. {\bf 35} 406

\bibitem{debroglie} 
  de Broglie L  1925 
  \APP {\bf 10} 22--128

\bibitem{smolin}
  Smolin L 2019 {\it Einstein's unfinished revolution: 
    the search for what lies beyond the quantum}
  (New York: Penguin Press) p 78

\bibitem{bloch} 
  Bloch F 1976 {\it Phys. Today}
  {\bf 29} 23--7

\bibitem{schrodinger}
  Schr\"odinger E 1926 
  \AP {\bf 384} 361--76

\bibitem{dicke}
  Dicke R H and Wittke J P 1960 {\it Introduction to Quantum
  Mechanics}
  (Reading, Massachusetts: Addison-Wesley) pp 23--37

\bibitem{gorard}
  Gorard J 2016 {\it Phys. Educ.} {\bf 51} 063003

\bibitem{chem}
  LibreTexts Chemistry Library, available online through 
  https://chem.libretexts.org

\bibitem{weinberg}
  Weinberg S 2015 {\it Lectures on Quantum Mechanics} 2nd ed
  (Cambridge: Cambridge University Press) pp 1--15

\bibitem{davisson}
  Davisson C and Germer L H 1927 
  {\it Nature} {\bf 119}, 558--60

\bibitem{zeilinger} 
  Zeilinger A, G\"ahler R, Shull C G, Treimer W and Mampe W 1988
  {\it Rev. Mod. Phys.} {\bf 60} 1067--73 

\bibitem{gasiorowicz}
  Gasiorowicz S 2003 {\it Quantum Physics} (New York: John Wiley \& Sons)
  pp 78--89

\bibitem{born}
  Born M 1926 \ZP {\bf 37} 863--7

\bibitem{sakurai}
  Sakurai J J 1985 {\it Modern Quantum Mechanics}
  Tuan S F ed 
  (New York: Addison Wesley Longman)
  p 489

\bibitem{griffiths}
  Griffiths D J 2005 {\it Introduction to Quantum Mechanics} 
  2nd ed (Upper Saddle River, NJ: Pearson Prentice Hall)
  p 1

\bibitem{auletta} Auletta G, Fortunato M, and  Parisi G 2009 
  {\it Quantum Mechanics} (Cambridge: Cambridge University Press)
\bibitem{banks} Banks T 2019 {\it Quantum Mechanics: An Introduction} 
   (Boca Raton: Taylor \& Francis Group)
\bibitem{binney} Binney J and Skinner D 2014 {\it The Physics of Quantum 
  Mechanics} (Oxford: Oxford University Press)
\bibitem{commins} Commins E D 2014 {\it Quantum Mechanics: an Experimentalist's Approach} (New York: Cambridge University Press)
\bibitem{mcintyre} McIntyre D H 2012 {\it Quantum Mechanics: a Paradigm Approach} 
  (San Francisco: Pearson Education)
\bibitem{basdevant} Basdevant J-L and Dalibard J 2002 {\it Quantum Mechanics} 
  (Berlin: Springer-Verlag)
\bibitem{townsend} Townsend J S 2012 {\it A Modern Approach to Quantum Mechanics} (Mill Valley: University Science Books)

\bibitem{fuller-malvern} 
  Fuller K D and Malvern D D 2010 {\it Challenge and Change: a History of the
  Nuffield A-Level Physics Project} (Reading: Institute of Education, 
  University of Reading) 
  available at  
  http://centaur.reading.ac.uk/7534/1/Malvern\underline{~}and\underline{~}Fuller.pdf

\end{thebibliography}
\end{document}